\documentclass[conference]{IEEEtran}
\IEEEoverridecommandlockouts
\usepackage{cite}
\usepackage{amsmath,amssymb,amsfonts}
\usepackage{algorithmic}
\usepackage{graphicx}
\usepackage{textcomp}
\usepackage{xcolor}

\usepackage[ruled,vlined]{algorithm2e}
\usepackage{hyperref}
\def\BibTeX{{\rm B\kern-.05em{\sc i\kern-.025em b}\kern-.08em
    T\kern-.1667em\lower.7ex\hbox{E}\kern-.125emX}}

\usepackage{booktabs} % For formal tables
\usepackage[utf8]{inputenc}
\usepackage[framemethod=default]{mdframed}
\usepackage{tcolorbox}
\usepackage{alltt}
\usepackage{soul}
\def\ttt{\texttt}

\definecolor{darkgreen}{RGB}{0,128,0}
\definecolor{darkblue}{RGB}{0,0,128}
\definecolor{darkred}{RGB}{192,0,0}

\definecolor{ltgray}{RGB}{192,192,192}
\definecolor{ltgold}{RGB}{248,230,180}

\definecolor{ltmagenta}{RGB}{255,204,255}
\definecolor{ltcyan}{RGB}{193,239,255}

\input{mathmode-spacing.tex}

\begin{document} 
\title{Test Suites as a Source of Training Data for Static Analysis Alert Classifiers}

\author{\IEEEauthorblockN{1\textsuperscript{st} Lori Flynn}
\IEEEauthorblockA{\textit{Software Engineering Institute} \\
\textit{Carnegie Mellon University}\\
Pittsburgh, USA \\
lflynn@sei.cmu.edu}
\and
\IEEEauthorblockN{2\textsuperscript{nd} William Snavely}
\IEEEauthorblockA{\textit{(former) Software Engineering Institute} \\
\textit{Carnegie Mellon University}\\
Pittsburgh, USA \\
will.snavely@gmail.com}
\and
\IEEEauthorblockN{3\textsuperscript{rd} Zachary Kurtz}
\IEEEauthorblockA{\textit{(former) Software Engineering Institute} \\
\textit{Carnegie Mellon University}\\
Pittsburgh, USA \\
zkurtz@pm.me}
}

\maketitle

\begin{abstract}
Flaw-finding static analysis tools typically generate large volumes of code flaw alerts including many false positives. To save on human effort to triage these alerts, a significant body of work attempts to use machine learning to classify and prioritize alerts. Identifying a useful set of training data, however, remains a fundamental challenge in developing such classifiers in many contexts. We propose using static analysis test suites (i.e., repositories of ``benchmark'' programs that are purpose-built to test coverage and precision of static analysis tools) as a novel source of training data. In a case study, we generated a large quantity of alerts by executing various static analyzers on the Juliet C/C++ test suite, and we automatically derived ground truth labels for these alerts by referencing the Juliet test suite metadata. Finally, we used this data to train classifiers to predict whether an alert is a false positive. Our classifiers obtained high precision (
\unskip%DO NOT EDIT DIRECTLY! Autoproduced in /zkurtz/classify/zach_push_results_to_paper.R
90.2\unskip
\%) and recall (
\unskip%DO NOT EDIT DIRECTLY! Autoproduced in /zkurtz/classify/zach_push_results_to_paper.R
88.2\unskip
\%) for a large number of code flaw types on a hold-out test set. This preliminary result suggests that pre-training classifiers on test suite data could help to jumpstart static analysis alert 
classification in data-limited contexts. 
\end{abstract}

\begin{IEEEkeywords}
static, analysis, alert, classification, rapid, precise, test suite, Juliet
\end{IEEEkeywords}

%\begin{document}
\section{Introduction}
Flaw-finding static analysis (FFSA) tools typically generate large volumes of code flaw alerts including many false positives. To save on human effort to triage these alerts, a significant body of work attempts to use machine learning to classify and prioritize alerts. Identifying a useful set of training data, however, remains a fundamental challenge in developing such classifiers in many contexts. We propose using static analysis test suites (i.e., repositories of ``benchmark'' programs that are purpose-built to test FFSA tools) as a novel source of training data for a wide range of ``conditions'' (types of code flaws). In a case study, we generated a large quantity of alerts by executing various FFSA tools on the Juliet test suite, and we automatically derived ground truth labels for these alerts by referencing the Juliet test suite metadata. Finally, we used this data to train classifiers to predict whether an alert is a false positive, and tested the classifiers on hold-out data.

This paper focuses on warnings from FFSA tools that look for security flaws. A checkerID is a unique string or regular expression in a tool's alerts for that flaw type. 
%We use regular expressions as a checkerID when no other string will work. %Compilers do much more than general flaw-finding static analysis, but many, like the Gnu Compiler Collection (GCC)~\cite{tool.gcc}, can output warnings about code flaws.
We use ``coding taxonomy'' to mean a named set of coding rules, weaknesses, standards, or guidelines (e.g., CERT coding rule~\cite{cert.standards} or Common Weakness Enumeration (CWE~\cite{mitre:cwe})). Each rule or weakness is considered a single condition. An FFSA tool's alert may be mapped to conditions in one or more coding taxonomies (e.g.,~\cite{mitre:cwecompat} lists many FFSA tools that provide CWE output). ``Alert fusion'' refers to the practice of unifying alert information from different tools which map to the same condition  in the same part of the code (e.g., same line of same file). Fusion may be imprecise, e.g., a particular CWE may occur in two different parts of the same line of code. 
\footnote{\footnotesize {\em AST 2021, May 20-21, 2021.}
This is the authors' version of the work. 
This material is based upon work funded and supported by the Department of Defense under Contract No. FA8702-15-D-0002 with Carnegie Mellon University for the operation of the Software Engineering Institute, a federally funded research and development center.
References herein to any specific commercial product, process, or service by trade name, trade mark, manufacturer, or otherwise, does not necessarily constitute or imply its endorsement, recommendation, or favoring by Carnegie Mellon University or its Software Engineering Institute. Carnegie Mellon\textregistered{} and CERT\textregistered{} are registered in the U.S. Patent and Trademark Office by Carnegie Mellon University. DM18-0144\par}

%% \footnote{\footnotesize Copyright 2021 IEEE. All Rights Reserved. This material is based upon work funded and supported by the Department of Defense under Contract No. FA8702-15-D-0002 with Carnegie Mellon University for the operation of the Software Engineering Institute, a federally funded research and development center.
%% References herein to any specific commercial product, process, or service by trade name, trade mark, manufacturer, or otherwise, does not necessarily constitute or imply its endorsement, recommendation, or favoring by Carnegie Mellon University or its Software Engineering Institute. Carnegie Mellon\textregistered{} and CERT\textregistered{} are registered in the U.S. Patent and Trademark Office by Carnegie Mellon University. DM18-0144\par} 

%\begin{table}[ht]
\begin{table}[!htbp]
\centering
\begin{tabular}{|l|p{5.3cm}|}
  \hline
{\bf Term} & {\bf Definition} \\ 
  \hline
SA tool & static analysis tool that analyzes code without running it\\ \hline
FFSA tool & flaw-finding static analysis tool\\ \hline
SA alert & static analysis alert (warning) about a particular type of flaw\\ \hline
Checker & analysis for a particular type of code flaw, by a particular FFSA tool %The checker may simply do syntactic analysis or do deep semantic analysis.
\\ \hline
CheckerID & checker name%, usually a unique string or regular expression in a tool's alerts
\\ \hline
Test suites & repositories of ``benchmark'' programs that are purpose-built to test FFSA tools\\ \hline
Condition & a constraint or property of validity with which code should comply. FFSA tools try to detect if code violates conditions.\\ \hline
Alert fusion & unifying alert information from different tools which map to the same condition and code location\\ \hline
Coding taxonomy & a named set of coding rules, weaknesses, standards, or guidelines\\ 
   \hline
\end{tabular}
\caption{Terminology} 
\label{table:terminology}
\end{table}

Table \ref{table:terminology} summarizes key terminology used in this paper.

\subsection{Related Work}\label{subsection:related.work}
Our work uses the Juliet C/C++ v1.2 test suite~\cite{nsacas:juliet.v.1.2}. The Juliet test suites~\cite{nsacas:juliet.v.1.2} provide example test programs with multiple subtypes of each CWE addressed (e.g., separate variant programs instantiating the flaw using an integer or a string), and for most of those subtypes also provides variant test programs involving variants of control, data, and/or type flow (e.g., a simple example with the flaw completely within a function, versus an example with control and data flow through multiple function calls and involving data pointers). Test suites are usually used to test and compare vulnerability analysis tools~\cite{delaitre2015evaluating, larsen2014state} or test system defenses~\cite{benameur2013minestrone}, but our work uses test suites in a new way: to rapidly build labeled data for SA alert classifier development. 

NIST provides over 600,000 cost-free, open-source test suite programs (including the Juliet test suite) in its Software Reference Dataset (SARD)~\cite{sard:test.suites}, along with metadata identifying each test's known flaw conditions and the flaw locations. Our method could be applied even more broadly with additional test suites including the others hosted by NIST.

Delaitre et al. tested security flaw-finding SA tools on the Juliet test suite (their research mostly focused on the Java tests) and on average those tools found about 20\% of weaknesses in basic test cases (no added complexity). They found that complex control flow or data flow constructs significantly reduced the tools' success rates, and identify flaws with highest and fewest findings across a set of anonymized FFSA tools~\cite{delaitre2013massive}. Since single FFSA tools have different coverage of code flaws (warning about some but not others)~\cite{delaitre2013massive, bessey2010few}, multiple FFSA tools may be used to find more code flaws~\cite{plakosh2014improving}. This approach, however, compounds the problem of generating too many alerts to deal with, including too many false positives. Our work uses multiple FFSA tools and addresses the issue of handling the alerts. Our work builds on the findings in~\cite{delaitre2013massive} by gathering data on FFSA tool performance for variant flows of control, type, and data for CWE sub-types and then using that for classifier development.

Pugh and Ayewah found a mean time of 117 seconds per SA alert review, from analyzing data from 282 Google engineers that made over 10,000 manual SA alert determinations~\cite{ayewah2010google}.
One example of insufficient labeled data as a barrier comes from our previous work with 3 large organizations that do software development, where lack of data covering more types of flaws resulted in classifier incorporation being impractical for them~\cite{flynn2017dod.case.study}. Even when an organization has large audit archives, if the auditors have not used a consistent set of audit rules and a well-defined auditing lexicon, the data may not be useful (and most organizations don't have a well-defined auditing lexicon and auditing rules)~\cite{svoboda2016}. Data-rich Google developed 85\% accurate classifier models predicting FindBugs false positives~\cite{ruthruff2008predicting}. Cross-project classifier prediction is an area of research developed to address insufficient labeled data for a code project. Our work is intended to rapidly develop a large quantity of labeled data archives using labeled alerts on test suite, creating classifiers to do cross-project defect prediction on production code alerts. 
Nam and Kim~\cite{nam2015clami} discuss issues related to cross-project defect prediction: some features in different projects may correlate to predictions in different ways, but the rationale for doing cross-project prediction is a hypothesis that some features are helpful for developing classifiers that work well across projects. They use the magnitude of metric values to do cross-project defect prediction. Their work focuses on determining a subset of labeled data and features within that data to use, while our work focuses on development of a labeled dataset covering many conditions that develops precise classifiers.
Zhang et al.~\cite{zhang2016cross} did cross-project prediction, showing improvement of classification results using a connectivity-based unsupervised classifier. A connectivity-based unsupervised classifier could possibly be used on data from the labeled audit archives produced from test suites and alerts from production code.
Jing et al.~\cite{jing2015heterogeneous} does cross-company defect prediction including cross-project prediction, using features that include the name of company that developed code for the line/function/file/class/program related to the alert. A labeled dataset from our system could be used in conjunction with their techniques.
 
Bessey et al.~\cite{bessey2010few} found many issues with using SA in practice, including tools ignoring constructs and thus producing many false positive alerts, users wrongly labeling diagnostics they find confusing as false, and user difficulty dealing with many alerts resulting in tools producing less (possibly-true) alerts about possible defects than they could. Beller et al. found that, in practice, few open-source projects have FFSA tools integrated closely with their workflows, and most of those projects do not mandate that a codebase should be warning-free~\cite{beller2016analyzing}. The goal of our work is to make it easier to use FFSA tools to effectively find and fix prioritized defects, including for open-source projects.

Heckman and Williams~\cite{heckman2011systematic} did an extensive survey of methods that classify and prioritize actionable alerts, detailing 21 peer-reviewed studies. Our method uses 5 of the approaches (alert type selection, contextual information, data fusion, machine learning, and mathematical and statistical models) discussed in the paper, and doesn't use the other 3 (dynamic detection, graph theory, and model checking). Our method uses 2 of the artifact characteristics categories (alert characteristics, code characteristics), and doesn't use the other 3 (source code repository metrics, bug database metrics, and dynamic analysis metrics). No previous work in their survey (nor in any research publication we have found since then) involves using test suites to automatically label data to create classifiers. 

The approach by Kremenek et al.~\cite{kremenek2004correlation} does not merge sets of alerts from different tools, so alerts from a tool are prioritized in a set according to that tool's ranking rather than individually. Alerts are ranked by correlating data per FFSA tool, using features from tools and from codebases~\cite{kremenek2003z}. Sets of alerts from different tools are ordered relatively (all alerts from one tool are prioritized below all alerts from another tool, never interspersed), with no alert fusion. The labeled data archives developed in our work could be used in combination with labeled production alerts in a similar adaptive heuristic.

Our own earlier research~\cite{flynn2016prioritizing, flynn2017dod.case.study} developed classifiers in many ways similar to the current work: labeled audit archives were used, with fused alerts from different tools that map to the same condition, line number, and filepath. However, that work was only able to develop accurate classifiers for 3 CERT C coding rules with single rule data, despite using a significant quantity of audit archives. Those audit archives include data from 8 years of CERT analysis on 26 codebases, plus new audit data provided by three collaborating organizations over the course of a year where the collaborators audited SA alerts for their own codebases using an auditing lexicon and auditing rules we developed~\cite{svoboda2016}. Our current work addresses that labeled-data quantity problem.

\section{New Method for Rapidly Generating Labeled Alert Archives}
\begin{figure*}
  \vspace{-2ex}
  \includegraphics[width=0.85\textwidth]{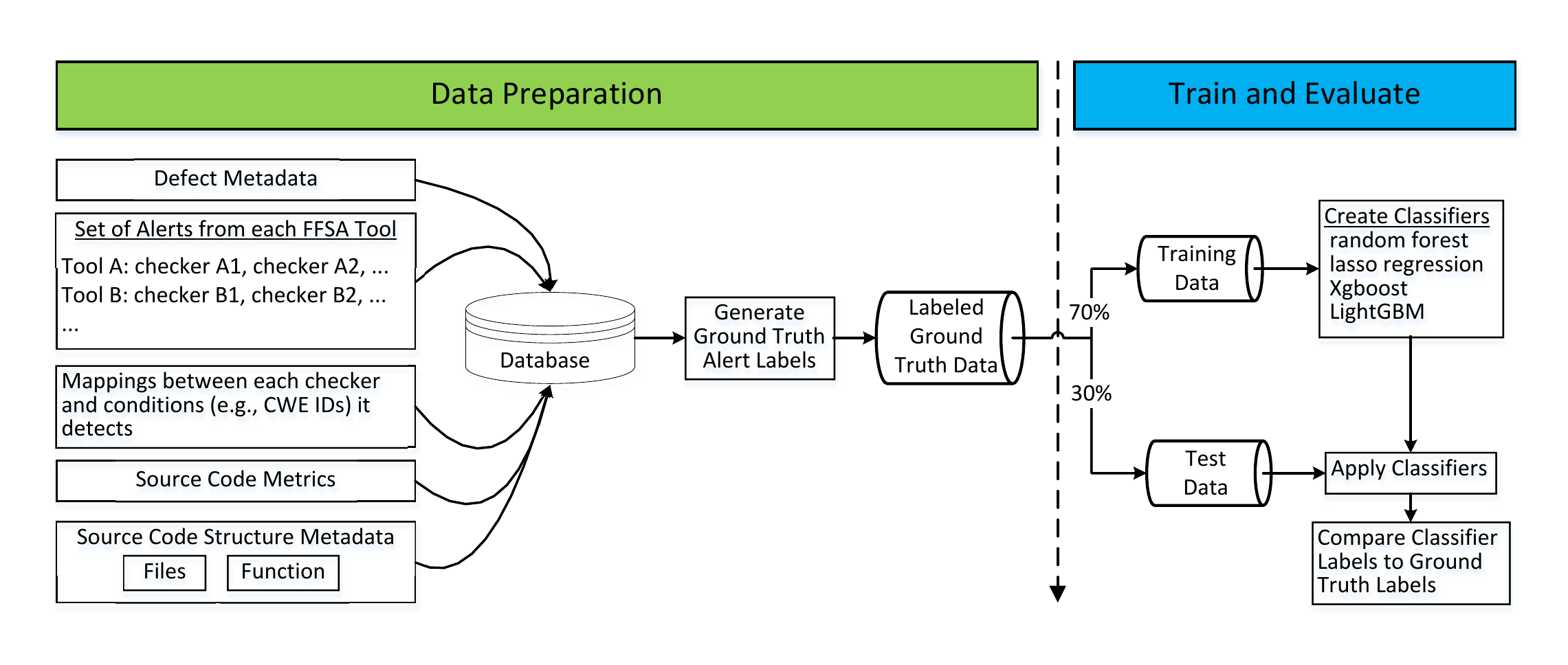}
  \vspace{-3ex}
  \caption{System for automated alert labeling and classifier training}
  \label{fig:system-for-automated-alert-labeling-for-classifiers}
\end{figure*}

To quickly generate a large quantity of labeled data for classifier development, we developed a software system that uses test suites, as shown in Fig.~\ref{fig:system-for-automated-alert-labeling-for-classifiers}. Data for classifier development uses results for each FFSA tool on variant flows of control, type, and data for CWE sub-types. Resultant classifier development potentially can use this information about individual FFSA tool precision (percent correct or true positive) in combination with type of flow complexity, CWE sub-type, and the set of FFSA tools that alert to more precisely predict if a given alert is true or false.

We ran 8 FFSA tools on the 61,387 tests (covering 118 CWEs) in the Juliet Test Suite for C/C++ v1.2~\cite{nsacas:juliet.v.1.2}, including popular proprietary tools and open-source tools. We developed scripts that parse the FFSA tool output and upload alert data to a database, including the checkerID plus the filepath and line number of the possible defect. Due to licensing restrictions, we cannot name proprietary tools we used nor their performance. The open-source FFSA tools we used are Cppcheck~\cite{tool.cppcheck}, Rosecheckers~\cite{tool.rosecheckers}, and GCC~\cite{tool.gcc}. GCC is a compiler, but like all compilers it statically analyzes the code, plus it outputs warnings (when run with parameters such as these: $-Wall -Wextra -Wpointer-arith  -Wstrict-prototypes -Wformat-security$) that can be mapped (as regular expressions) to code flaw taxonomies \cite{cert:secure.coding.wiki.analyzers.gcc}.

%\vspace{-4ex}

We generated additional features for the classifiers to train on by running three code metrics tools on the test suite: (Lizard~\cite{tool.lizard}, CCSM~\cite{tool.ccsm}, and a proprietary tool (anonymous)). All the tools provide counts of significant lines of code, complexity, and cohesion. Lizard provided the fewest metrics and CCSM provided the most. Our database correlates each alert with applicable code metrics.
%Relationships between collections in the Mongo database our system creates are shown in Fig.~\ref{fig:relationships-between-collections-in-mongo-db}.

\subsection{Deriving Ground Truth Alert Verdicts}\label{subsection:alert.determination}
The Juliet Test Suite contains two kinds of metadata that are relevant for determining the validity of alerts: 
\begin{itemize}
\item {\bf a manifest file}: This is an \ttt{XML} file that provides precise flaw information including line number, CWE, and filepath. 
\item {\bf function names}: Documentation for the test suite says that if the function name includes the string \ttt{GOOD} then the particular CWE does not occur in it, but if it includes the string \ttt{BAD} then the CWE does occur in the function. We gathered information about filepath and line numbers covered by each function name that contains \ttt{GOOD} or \ttt{BAD}, as well as the CWE indicated (usually by filename).
\end{itemize}

Note that both the manifest file and the function names provide only CWE-specific flaw information.  In general, a line of code marked BAD for one code flaw type could be flawless with respect to all other code flaw types. Thus, we can use the metadata flaw information to determine the validity of an alert only when we can establish that the alert's checkerID is for a flaw of the same type as a flaw referenced in the metadata. The test suite metadata does not identify every CWE weakness in the Juliet code nor all locations of the CWE weaknesses, so an alert that doesn't map to the test suite metadata cannot be automatically labeled using the metadata. In other words, if an alert's CWE doesn't match the test suite metadata's CWE, the metadata can't be used to label the alert true or false.

Publicly-available mappings between checkerIDs and CWEs are available for many of the FFSA tools that we tested. We fused alerts from this set of tools, producing a set of fused alerts with known CWE designations. We then determined verdicts (i.e. classifier ground truth labels) for each fused alert as follows:
\begin{itemize}
\item If the manifest includes a record of a flaw for the same filepath, line number, and CWE as the alert, then set \allowbreak \ttt{verdict}$=$\allowbreak\ttt{True}, indicating that the alert is a true positive. 
\item If the defect alert matches {\em any} line within a function name with \ttt{GOOD} {\em and} the alert's CWE matches the CWE associated with the function, then set \ttt{verdict}$=$\ttt{False}, indicating a that the alert is a false positive.
\end{itemize}

Applying the above procedure resulted in 
%DO NOT EDIT DIRECTLY! Autoproduced in /zkurtz/classify/ophelia/publish/compile_results_for_icse2019.R
36,968\unskip TP fused alerts,
%DO NOT EDIT DIRECTLY! Autoproduced in /zkurtz/classify/ophelia/publish/compile_results_for_icse2019.R
84,269\unskip FP fused alerts, and 
%DO NOT EDIT DIRECTLY! Autoproduced in /zkurtz/classify/ophelia/publish/compile_results_for_icse2019.R
\unskip fused alerts with no determination. Note that the procedure defined above conservatively refrains from assigning a verdict for an alert when (a) the alert falls within a function labeled BAD but (b) the metadata flaw line number does not match the line number of the alert.

Automatically generating this volume of alert verdicts is potentially a nontrivial cost savings. As per Section~\ref{subsection:related.work}), manually auditing alerts typically takes on the order of 
%DO NOT EDIT DIRECTLY! Autoproduced in /zkurtz/classify/ophelia/publish/compile_results_for_icse2019.R
117\unskip seconds per alert. Manually generating our
%DO NOT EDIT DIRECTLY! Autoproduced in /zkurtz/classify/ophelia/publish/compile_results_for_icse2019.R
121,237\unskip alert verdicts at this rate might have taken
%DO NOT EDIT DIRECTLY! Autoproduced in /zkurtz/classify/ophelia/publish/compile_results_for_icse2019.R
3,940\unskip hours. 
However, manual auditing and natural code would likely not cover many of the conditions in the Juliet test suite, since many conditions rarely exist in natural code. Even if FFSA tools were run on a large number of natural codebases (which would take a lot of time and computation) and alerts were found for many conditions, it might take many manual audits before both true and false determinations could be made. Realistically, it would take an enormous amount of manual auditing time (and money to pay for it), to develop that much data.

In comparison, incorporating a new FFSA tool into our system takes approximately 24 hours of developer effort, to code a parser for the FFSA tool output (unless it uses the new SARIF standard output format~/cite{sarif}) and otherwise integrate into the system. After the FFSA tool is run on the test suite, the script the developer wrote is run on the output, and then labeling the alerts automatically and rapidly happens. 
Adding a new test suite to our system similarly involves using standard interfaces to adapt the system to integrate another test suite. However, adding a new test suite takes more manual effort because each test suite has new unique compile and test requirements, and for non-SARD test suites the metadata format must be adapted to the SARD metadata format.
Considering the effort required to add more FFSA tools and test suites, the new automated system clearly scales to generate much more labeled data at much lower cost than manual analysis.

\subsection{Speculative Mapping Method}\label{subsection:spec_mapping}
For FFSA tools that did not have publicly-available checker mappings to CWEs, we developed a \emph{speculative mapping} method to estimate these mappings. Since FFSA tools often find different sets of 
defects, we hoped to obtain greater CWE coverage without the cost of manual mapping between the FFSA tools' checkers and CWEs. 

Our speculative mapping method treats a co-occurrence of a test suite metadata CWE designation for a line of code and an alert on that same line of code as \emph{evidence} that the alert's checkerID was designed to detect the kind of flaw indicated by that CWE. Such a co-occurence is a \emph{checker-CWE match}.

We summarize this evidence by counting all checker-CWE matches. That is, for every checker-CWE pair, we count how many times an alert issued by that checker falls on the same line as a code flaw of that CWE. A checker may match multiple CWEs, however, so it's not always obvious which CWE a checker is the \emph{best} match for.

We define two checker-CWE \emph{match rate percentages} to guide the speculative mapping of checkerIDs to CWEs. Let $m_{ij}$ denote the count of checker-CWE matches between checker $i$ and CWE $j$; let $m_i = \sum_j m_{ij}$, the total count of all checker-CWE matches involving checker $i$; and let $m_j = \sum_i m_{ij}$, the total count of all checker-CWE matches (for the tool being speculatively mapped) involving CWE $j$. Finally, we define the
$$ \mbox{forward match percentage} = 100 \frac{m_{ij}}{m_i} $$
and the 
$$ \mbox{backward match percentage} = 100 \frac{m_{ij}}{m_j}. $$
Conceptually, these match rate percentages are alternative ways to measure the relative frequencies by which alerts from each checker fall on the lines of particular conditions (CWEs) that are recorded in the test suite manifest.

For each checker, we establish a preliminary assignment of that checker to the CWE that has the greatest forward (backward) match rate percentage with that checker. Finally, we fix a threshold greater than $0$. All preliminary assignments for which the forward (backward) match rate exceeds the threshold become \emph{speculative mappings}. (Section \ref{subsection:speculative.mapping.results} details the specific thresholds used.) 

Proceeding with the assumption that the speculative mappings are correct allows us to increase the number of alerts with ground truth verdicts. Every alert from a speculatively-mapped checker has an associated CWE (via the speculative mapping), and we assign verdicts for such alerts just as described in Section \ref{subsection:alert.determination}.

\subsubsection{Errors in Speculative Mappings}
Some speculative mappings are wrong. For example, the unmapped tool's checker may be alerting about a type of code flaw on that code line that the test suite manifest does not mention.
Other mappings are only partially correct. That is the case when a CWE and a tool checker do not have an $EQUALS$ relationship, but instead have a $SUBSET OF$, $SUPERSET OF$, or $PARTIALLY OVERLAPPING$ relationship.
Some mappings will be missed by speculative mapping. The unmapped tool might detect the code flaw specified in the test suite manifest, but assign the flaw location to code location $X$ while the manifest assigns it to code location $Y$. Fault and failure location may differ. For example, if a failure is caused by an out of bound array index, the tool and manifest might locate the fault differently. One could define it in the very line of the access or in a function that did not sanitize before that.
  
\subsubsection{Test Mappings Rationale and Alternatives}
We tested forward and backward speculative matching with varying thresholds, to gather empirical data related to checker to CWE mappings. We tried to identify generally-useful thresholds or mapping directions. If the actual relationship is $EQUALS$ then we would expect close to a $100\%$ match. If the actual relationship is $SUBSET OF$ or $SUPERSET OF$ then we would expect the forward and backward speculative matching results to be different. If the relationship is $PARTIALLY OVERLAPPING$ the forward and backward speculative matching results could be similar or different. Alternative formulas not tested in this work might produce useful results, such as: $combined\_match\_percentage = 0.5*forward + 0.5*backward$.

\subsection{Automated Speculative Mapping with Manual Verification}\label{subsection:spec_mapping}
We identified an alternative to using the automated speculative mappings alone. In this alternative, we start with speculative mappings and then perform manual verification of the mapping. There are approximately 700 CWE and static analysis tools may have hundreds of checker IDs. Validating $70,000$ possible checker mappings (in this example, for a tool with only 100 checker IDs) would be very time-consuming, as it requires a precise understanding of both the checker definition (which takes on average $time(D\_checker)$) and the CWE definition (which takes on average $time(D\_CWE)$), and after that determination of the relationship between checker and CWE (which takes on average $time(Relationship)$) for a total time in our example of $70,000*time(D\_CWE)*time(D\_checker)*time(Relationship)$. However, speculative mapping reduces the candidate mappings to a far smaller number, with each candidate mapping by definition having evidence supporting the possible mapping. Three different speculative CWE mappings per each tool checker ID is far more than we actually observed. Using that overestimate, for our example the amount of time required to do such mapping would be $300*time(D\_CWE)*time(D\_checker)*time(Relationship)$, which is a factor of $233$ less than $70,000$.

\section{Building and testing classifiers}\label{section:testing}

The classifier task is to determine whether a fused alert is a True Positive (TP) or a False Positive (FP) on the basis of various code metrics and other features associated with each alert. Section \ref{section:train.test} explains how we divided the labeled alert archives into a training set and a test set. Section \ref{subsection:classifiers} introduces the classifiers. Section \ref{subsection:basic.alerts} describes the performance of the classifiers for fused alerts excluding all speculatively mapped alerts and analyzes the importance of specific features. Section \ref{subsection:speculative.mapping.results} shows how overall performance differs after including speculatively mapped alerts in the training data. 

\subsection{Training data versus test data}\label{section:train.test}

We defined a non-speculative training data set and numerous speculative training data sets based on the inclusion or exclusion of alerts with speculatively-mapped checkerIDs. The hold-out test set, by contrast, involved only alerts with non-speculative mappings to ensure a consistent comparison between a classifier that is alternatively trained on the two training sets.

To be precise about the test train splits, define the following sets of raw (pre-fused) alerts and fused alerts:
\begin{itemize}
\item $A_{mapped}$: the raw alerts with checkerIDs that have known CWE mappings.
\item $A(T,d)$: the raw alerts in $A_{mapped}$ together with the alerts whose checkerID that is speculatively mapped to a CWE above the threshold T on the match percentage in direction $d \in \{$forward, backward$\}$, as per Section \ref{subsection:spec_mapping}.
\item $A_{pure}$: the raw alerts that do not share a line with any other alert that has a speculatively-mapped checkerID (for every threshold and match percentage direction).
\item $AF_{mapped}$: the fused alerts derived from $A_{mapped}$.
\item $AF(T, d)$: the fused alerts derived from $A(T,d)$.
\item $AF_{pure}$: the fused alerts derived from $A_{pure}$.
\item $AF_{test}$: a stratified (on \ttt{verdict} and CWE) random sample of the fused alerts in $AF_{pure}$. This represents a ``pure" test set not intertwined with speculative mappings.
\item $AF_{non-speculative}$: the fused alerts in $AF_{mapped}$ excluding those in $AF_{test}$.
\item $AF_{speculative}(T, d)$: the fused alerts in $AF(T, d)$, excluding those in $AF_{test}$. 
\end{itemize}

% We establish a ``pure" test set that not intertwined with speculative mappings, we identify the set $A$ of alerts that do not share a line with any other alert that has an unmapped checkerID.  HOWEVER See zkurtz/rcr/R/standard_analysis::get_test_fusion_ids
% #' OPEN QUESTION: why not exclude only fusion_ids for which ALL alerts
% #'     are speculative? The addition of a speculatively mapped alert to an
% #'     established fusion ID corrupts it how exactly? ...

To summarize, we started by defining our single test set $AF_{test}$ which included %DO NOT EDIT DIRECTLY! Autoproduced in /zkurtz/classify/ophelia/publish/compile_results_for_icse2019.R
36,445\unskip fused alerts, or about 30\% of the fused alerts excluding speculative mappings. We then defined the training set $AF_{non-speculative}$ to include all the remaining fused alerts, excluding speculative mappings. Finally, we defined a series of training sets $AF(T, d)$ that include all of the alerts in $AF_{non-speculative}$ but also include varying subsets of the alerts with speculatively-mapped checkerIDs.

Table \ref{table:n_fused_alerts_at_thresholds} shows the number of fused alerts in the training set $AF_{speculative}(T, d)$ for each combination of threshold \\
$T \in \{0\% , 5\%, 25\%, 50\%, 75\%, 100\%\}$ and match direction $d$. At the most permissive setting (with a threshold of $0$), the number of fused alerts in the training data is 
%DO NOT EDIT DIRECTLY! Autoproduced in /zkurtz/classify/speculative_mappings/rc224_comparisons.R
464,887\unskip, approximately twice the 
%DO NOT EDIT DIRECTLY! Autoproduced in /zkurtz/classify/ophelia/publish/compile_results_for_icse2019.R
121,237\unskip\space fused alerts in the main training set.

% latex table generated in R 3.5.0 by xtable 1.8-2 package
% Tue Sep 25 13:29:46 2018
\begin{table*}[ht]
\centering
\begin{tabular}{lrrrrrr}
  \hline
Match direction & 0\% & 5\% & 25\% & 50\% & 75\% & 100\% \\ 
  \hline
backward & 464887 & 308038 & 169370 & 149926 & 137128 & 134374 \\ 
  forward & 464887 & 426927 & 296259 & 222108 & 193622 & 134374 \\ 
   \hline
\end{tabular}
\caption{Number of fused alerts in the training data for each speculative mappings threshold} 
\label{table:n_fused_alerts_at_thresholds}
\end{table*}

\subsection{Classifiers and performance metrics}
\label{subsection:classifiers}
We used the \verb|R| statistical programming language to run four different classifiers: LightGBM, XGBoost, the \verb|H2O.ai| implementation of random forests, and the \verb|glmnet| implementation of lasso-regularized logistic regression. All of these except for the lasso regression are based on decision trees. LightGBM and XGBoost won acclaim on Kaggle~\cite{kaggle} and are both examples of regularized Gradient Boosting Machines. Random forest and lasso regression are both generic algorithms; we used the the random forest implementation from \verb|h2o.ai| and the lasso implementation in the \verb|glmnet| package.

Our purpose for trying several classifiers is to identify at least one one that generates highly accurate predictions on a hold-out test set. In particular, inference regarding the importance of individual features (including multicollinearity-induced variance in feature effects) was not a primary focus. Although we did some experimentation to choose reasonable hyperparameters for each classifier, we largely accepted default settings, treating each algorithm as a black-box. All four of these algorithms are sophisticated in the sense that they tend to perform well on nearly arbitrary sets of features.

For each of these classifiers, a prediction for an alerts is a number between $0$ and $1$ that represents an estimated probability that the alert has \ttt{verdict}$=$\allowbreak\ttt{True}. We round these predictions to generate binary classification output as needed to compute metrics like precision and recall.

Our primary metric for classifier performance is the area under the receiver operating characteristic curve (AUROC). The AUROC is useful for comparing the overall performance of classifiers when the classifier output is probabilistic. The AUROC gives partial credit to probabilistic predictions: When the true label is $1$, a prediction of $0.4$ scores much better than a prediction of $0$, for example. The AUROC penalizes false positive probability mass and false negative probability mass equally. 

In addition to the AUROC, we compute the precision, recall, and accuracy, since these standard metrics are relatively easy to interpret. The \verb|precision| is the fraction of predicted TP alerts that were indeed TP, and the \verb|recall| is the fraction of TP alerts that were successfully predicted to be TP. Note that precision is meaningless when the classifier does not classify any issues as TP, and so precision values for certain subsets of the test set are missing. Similarly, \verb|recall| is meaningless when the labeled alerts contain no TP alerts, as there is nothing to be ``recalled''. 

\subsection{Results without speculative mappings} \label{subsection:basic.alerts}

This section summarizes performance of the classifiers on the test set $AF_{test}$ after training on the main training data $AF_{non-speculative}$ (as per Section \ref{section:train.test}).

Table~\ref{table:per_classifier_summary} summarizes the average performance of the different classifiers over the entire test set. Table~\ref{table:lightgbm_for_cwes} summarizes the performance of LightGBM, our best classifier, on all CWE IDs for which at least one test data point was available. Note that that \verb|test count| is the number of fused alerts available for testing for each CWE ID, and \verb|TP rate| is the fraction of the testing alerts that were TP. 

Table~\ref{table:lightgbm_for_cert_rules} shows the same performance statistics on groups of alerts associated with selected CERT rules. For each CERT secure coding rule, we identified the set of relevant CWE test programs based on previously-established mappings between the taxonomies, documentation about the test suite, and manual inspection of the test suite programs. We subsequently identified the set of fused alerts applicable to each CERT rule via their CWE designations. These sets are not mutually disjoint in general; the CWE of each fused alert can correspond to no CERT rule, one CERT rule, or multiple CERT rules. Thus, the values of \verb|test count| need not correspond to the counts in Table~\ref{table:lightgbm_for_cwes}. %For each CERT rule, a we selected a fraction of its fused alerts as part of the classifier evaluation test set, and these were the basis for the classifier performance results in Table \ref{table:lightgbm_for_cert_rules}. 

A cursory examination of \verb|accuracy| in Tables~\ref{table:lightgbm_for_cwes} and \ref{table:lightgbm_for_cert_rules} suggests that the classifier is often accurate across many different kinds of rules. Some of these results stand out. The first CWE (457) and second CERT Rule (\verb|ARR30-C|) both have very low TP rates; such low TP rates are the source of a great deal of wasted triaging effort. Our classifier, however, performed at nearly 100\% accuracy on these sets of alerts, correctly classifying them as FP.

There are limitations, however, to how broadly we can interpret this success. Juliet Test Suite is not a representative sample of code in general. The kinds of code artifacts that trigger alerts from various checkers on the test suite could differ significantly from other code bases, degrading the performance of our classifier outside of the test suite. 
% latex table generated in R 3.5.0 by xtable 1.8-2 package
% Thu Sep 20 17:16:43 2018
\begin{table}[ht]
\centering
\begin{tabular}{lrrrr}
  \hline
Classifier & Accuracy & Precision & Recall & AUPRC \\ 
  \hline
rf & 0.947 & 0.898 & 0.902 & 0.979 \\ 
  lasso & 0.880 & 0.891 & 0.628 & 0.833 \\ 
  xgboost & 0.949 & 0.922 & 0.882 & 0.974 \\ 
  lightgbm & 0.958 & 0.917 & 0.928 & 0.985 \\ 
   \hline
\end{tabular}
\caption{Average performance of different classifiers} 
\label{table:per_classifier_summary}
\end{table}

% latex table generated in R 3.5.0 by xtable 1.8-2 package
% Thu Sep 20 17:17:03 2018
\begin{table}[ht]
\centering
\begin{tabular}{lrrrrr}
  \hline
CWE-ID & test count & TP rate & precision & recall & accuracy \\ 
  \hline
824 & 3239 & 0.09 & 1.00 & 1.00 & 1.00 \\ 
  457 & 2906 & 0.06 & 1.00 & 1.00 & 1.00 \\ 
  681 & 2871 & 0.25 & 0.85 & 0.87 & 0.93 \\ 
  665 & 2467 & 0.03 & 1.00 & 1.00 & 1.00 \\ 
  908 & 2451 & 0.03 & 1.00 & 1.00 & 1.00 \\ 
  758 & 2317 & 0.00 & 1.00 & 1.00 & 1.00 \\ 
  195 & 1485 & 0.22 & 0.73 & 0.95 & 0.91 \\ 
  194 & 1286 & 0.25 & 0.72 & 0.95 & 0.89 \\ 
  196 & 1265 & 0.24 & 0.83 & 0.80 & 0.92 \\ 
  676 & 994 & 0.44 & 0.82 & 0.82 & 0.84 \\ 
  426 & 982 & 0.43 & 0.84 & 0.74 & 0.83 \\ 
  78 & 982 & 0.43 & 0.82 & 0.80 & 0.83 \\ 
  704 & 866 & 0.40 & 0.82 & 0.90 & 0.88 \\ 
  253 & 853 & 0.27 & 1.00 & 1.00 & 1.00 \\ 
  762 & 850 & 0.67 & 1.00 & 1.00 & 1.00 \\ 
  404 & 734 & 0.56 & 1.00 & 0.99 & 1.00 \\ 
  761 & 698 & 0.60 & 1.00 & 1.00 & 1.00 \\ 
  197 & 693 & 0.44 & 0.94 & 0.73 & 0.86 \\ 
  401 & 663 & 0.53 & 1.00 & 1.00 & 1.00 \\ 
  680 & 618 & 0.17 & 0.73 & 0.93 & 0.93 \\ 
  476 & 513 & 0.23 & 0.99 & 0.99 & 1.00 \\ 
  690 & 501 & 0.17 & 0.99 & 0.99 & 1.00 \\ 
  190 & 461 & 0.16 & 1.00 & 1.00 & 1.00 \\ 
  20 & 446 & 0.64 & 1.00 & 1.00 & 1.00 \\ 
  188 & 424 & 0.33 & 0.83 & 0.91 & 0.91 \\ 
  672 & 417 & 0.60 & 1.00 & 1.00 & 1.00 \\ 
  134 & 381 & 0.44 & 0.97 & 0.79 & 0.90 \\ 
  775 & 376 & 0.18 & 0.98 & 0.97 & 0.99 \\ 
  191 & 361 & 0.29 & 0.77 & 0.79 & 0.87 \\ 
  129 & 336 & 0.49 & 1.00 & 0.98 & 0.99 \\ 
  590 & 300 & 1.00 & 1.00 & 1.00 & 1.00 \\ 
  628 & 279 & 0.88 & 0.98 & 0.98 & 0.97 \\ 
  369 & 234 & 0.47 & 0.96 & 0.97 & 0.97 \\ 
  606 & 232 & 0.31 & 1.00 & 1.00 & 1.00 \\ 
  119 & 225 & 0.73 & 1.00 & 1.00 & 1.00 \\ 
  252 & 167 & 0.74 & 1.00 & 1.00 & 1.00 \\ 
  456 & 148 & 0.54 & 1.00 & 1.00 & 1.00 \\ 
  909 & 139 & 0.49 & 1.00 & 1.00 & 1.00 \\ 
  125 & 135 & 0.56 & 1.00 & 1.00 & 1.00 \\ 
  121 & 118 & 0.82 & 0.93 & 0.95 & 0.90 \\ 
  122 & 116 & 0.83 & 0.98 & 0.90 & 0.90 \\ 
  843 & 115 & 0.23 & 0.81 & 0.78 & 0.90 \\ 
  126 &  71 & 0.83 & 0.93 & 0.97 & 0.92 \\ 
  563 &  65 & 0.17 & 1.00 & 1.00 & 1.00 \\ 
  377 &  64 & 0.00 &  &  & 1.00 \\ 
  415 &  37 & 0.73 & 1.00 & 0.96 & 0.97 \\ 
  468 &  36 & 0.36 & 1.00 & 1.00 & 1.00 \\ 
  469 &  35 & 0.37 & 1.00 & 1.00 & 1.00 \\ 
  398 &  34 & 0.38 & 1.00 & 1.00 & 1.00 \\ 
  192 &  33 & 0.00 &  &  & 1.00 \\ 
  480 &  32 & 0.97 & 1.00 & 1.00 & 1.00 \\ 
  783 &  32 & 0.97 & 1.00 & 1.00 & 1.00 \\ 
  127 &  27 & 0.96 & 0.96 & 0.96 & 0.93 \\ 
  327 &  25 & 1.00 & 1.00 & 1.00 & 1.00 \\ 
  416 &  23 & 0.56 & 1.00 & 1.00 & 1.00 \\ 
  688 &  22 & 0.00 &  &  & 1.00 \\ 
  569 &  20 & 0.95 & 1.00 & 1.00 & 1.00 \\ 
  787 &  20 & 0.60 & 1.00 & 1.00 & 1.00 \\ 
  120 &  19 & 1.00 & 1.00 & 1.00 & 1.00 \\ 
  131 &  19 & 1.00 & 1.00 & 1.00 & 1.00 \\ 
  170 &  19 & 1.00 & 1.00 & 1.00 & 1.00 \\ 
  328 &  19 & 1.00 & 1.00 & 1.00 & 1.00 \\ 
  467 &  19 & 1.00 & 1.00 & 1.00 & 1.00 \\ 
  587 &  18 & 0.39 & 1.00 & 1.00 & 1.00 \\ 
  570 &  16 & 1.00 & 1.00 & 1.00 & 1.00 \\ 
  367 &  13 & 1.00 & 1.00 & 1.00 & 1.00 \\ 
  465 &  13 & 1.00 & 1.00 & 1.00 & 1.00 \\ 
  597 &  13 & 1.00 & 1.00 & 1.00 & 1.00 \\ 
  483 &  11 & 0.00 &  &  & 1.00 \\
  338 &   7 & 1.00 & 1.00 & 1.00 & 1.00 \\ 
  478 &   7 & 1.00 & 1.00 & 1.00 & 1.00 \\ 
  482 &   7 & 1.00 & 1.00 & 1.00 & 1.00 \\ 
  484 &   7 & 1.00 & 1.00 & 1.00 & 1.00 \\ 
  123 &   6 & 0.83 & 0.83 & 1.00 & 0.83 \\ 
  561 &   5 & 1.00 & 1.00 & 1.00 & 1.00 \\ 
  571 &   4 & 1.00 & 1.00 & 1.00 & 1.00 \\ 
  466 &   1 & 1.00 & 1.00 & 1.00 & 1.00 \\ 
  562 &   1 & 1.00 &  & 0.00 & 0.00 \\ 
  835 &   1 & 1.00 & 1.00 & 1.00 & 1.00 \\ 
   \hline
\end{tabular}
\caption{Performance of LightGBM on test sets composed of meta alerts for specific CWE IDs.} 
\label{table:lightgbm_for_cwes}
\end{table}

% latex table generated in R 3.5.0 by xtable 1.8-2 package
% Thu Sep 20 17:17:17 2018
\begin{table}[ht]
\centering
\begin{tabular}{llrrrr}
  \hline
CERT rule & test count & TP rate & precision & recall & accuracy \\ 
  \hline
ARR30-C & 2123 & 0.30 & 0.81 & 0.84 & 0.89 \\ 
  ARR36-C & 2015 & 0.06 & 1.00 & 1.00 & 1.00 \\ 
  ARR38-C & 973 & 0.16 & 0.84 & 0.85 & 0.95 \\ 
  ARR39-C & 631 & 0.03 & 1.00 & 0.94 & 1.00 \\ 
  CON43-C & 615 & 0.12 & 0.81 & 0.85 & 0.96 \\ 
  DCL30-C & 372 & 0.42 & 0.97 & 0.78 & 0.90 \\ 
  ENV33-C & 371 & 0.16 & 0.98 & 0.97 & 0.99 \\ 
  ERR33-C & 271 & 0.07 & 1.00 & 1.00 & 1.00 \\ 
  ERR34-C & 259 & 1.00 & 1.00 & 1.00 & 1.00 \\ 
  EXP33-C & 240 & 0.00 &  &  & 1.00 \\ 
  EXP34-C & 234 & 0.47 & 0.96 & 0.97 & 0.97 \\ 
  EXP37-C & 197 & 1.00 & 1.00 & 1.00 & 1.00 \\ 
  EXP45-C & 186 & 0.40 & 0.81 & 0.95 & 0.89 \\ 
  EXP46-C & 140 & 0.12 & 1.00 & 1.00 & 1.00 \\ 
  FIO30-C & 106 & 0.85 & 0.97 & 0.92 & 0.91 \\ 
  FIO42-C & 93 & 1.00 & 1.00 & 1.00 & 1.00 \\ 
  FIO47-C & 57 & 0.75 & 0.93 & 0.93 & 0.89 \\ 
  FLP32-C & 48 & 0.31 & 0.93 & 0.93 & 0.96 \\ 
  FLP34-C & 35 & 0.37 & 1.00 & 1.00 & 1.00 \\ 
  INT30-C & 31 & 0.26 & 1.00 & 1.00 & 1.00 \\ 
  INT31-C & 18 & 0.39 & 1.00 & 1.00 & 1.00 \\ 
  INT32-C & 13 & 1.00 & 1.00 & 1.00 & 1.00 \\ 
  INT33-C & 7 & 1.00 & 1.00 & 1.00 & 1.00 \\ 
  INT36-C & 1 & 1.00 &  & 0.00 & 0.00 \\ 
  MEM30-C & 1 & 1.00 & 1.00 & 1.00 & 1.00 \\ 
   \hline
\end{tabular}
\caption{Performance of LightGBM on test sets composed of fused alerts that correspond to specific CERT rules.} 
\label{table:lightgbm_for_cert_rules}
\end{table}

\begin{figure*}
  \vspace{-2ex}
  \includegraphics[width=0.95\textwidth]{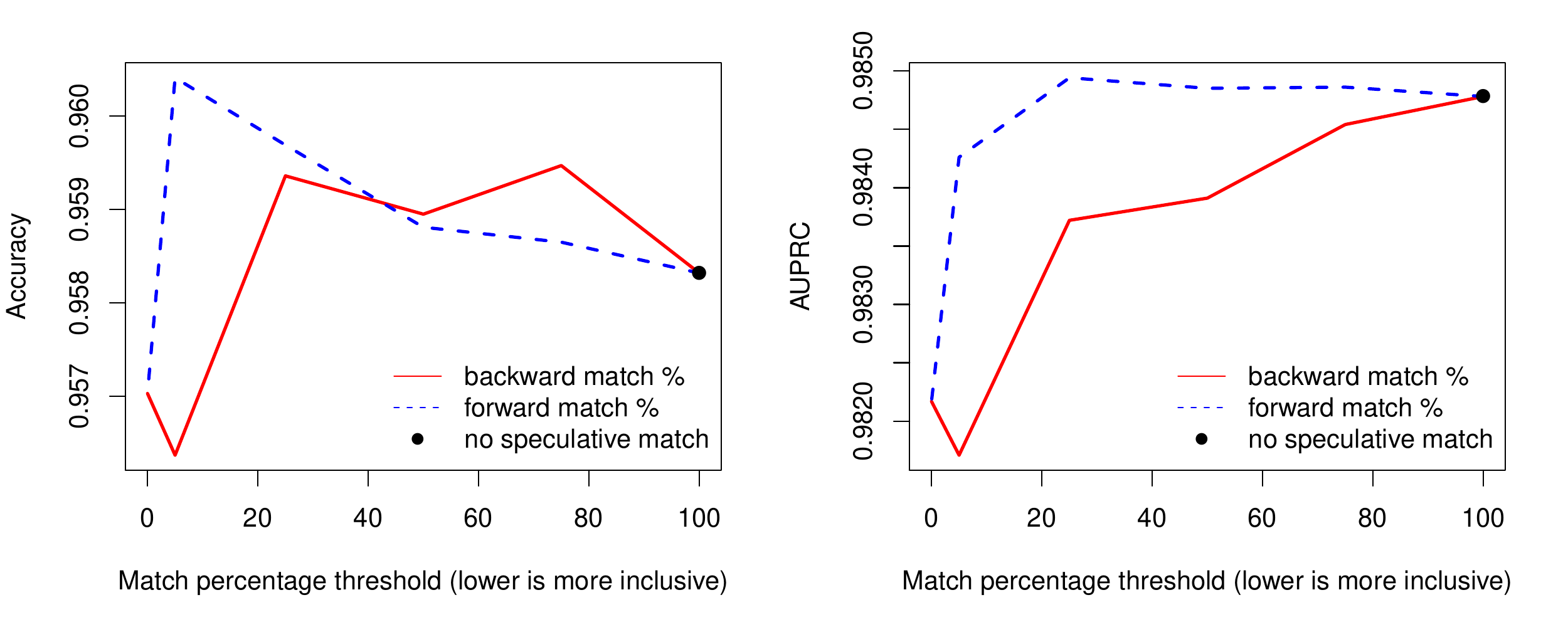}
  \vspace{-4ex}
  \caption{Assessing the value of training a classifier on speculatively mapped alerts}
  \label{fig:SM_training_value}
\end{figure*}

%\begin{table}[ht]
\begin{table}[htbp]
\centering
\begin{tabular}{lrr}
  \hline
Rank & Feature & Importance \\ 
  \hline
  $1$ &  CCSM \ttt{FUNC\_CALLED\_BY\_LOCAL} & $0.423$ \\ 
  $2$ &  checkerID & $0.294$ \\ 
  $3$ &  CCSM \ttt{RAW\_KW\_LONG\_CNT}& $0.047$ \\ 
  $4$ &  Tool\_X \ttt{Metric 1} & $0.031$ \\ 
  $5$ &  Tool\_X \ttt{Metric 2} & $0.026$ \\ 
  $6$ & CWE & $0.025$ \\ 
  $7$ & Lizard \ttt{Parameters} & $0.015$ \\ 
  $8$ &  Tool\_X Metric 3 & $0.012$ \\ 
  $9$ &  Lizard \ttt{Number\_of\_Tokens} & $0.008$ \\ 
  $10$ &  Tool\_X \ttt{Metric 4} & $0.006$ \\ 
   \hline
\end{tabular}
\caption{Feature Importance in LightGBM: Top 10.} 
\label{table:feature_importance}
\end{table}

%\subsection{Feature Importance}
We calculated the top 10 features in terms of the total gain of each feature's splits in the underlying LightGBM trees, shown in Table~\ref{table:feature_importance} (metrics from a proprietary tool anonymized). This importance ranking is not guaranteed to identify all significant features, and the precise ordering of features is not very meaningful (in the presence of multicollinearity, for example). Such a ranking, however, is useful for identifying at least a sample of the moderate-to-highly-significant features. We use this to provide an example of the patterns that the classifier detects. 

\begin{figure}
  \includegraphics[width=0.47\textwidth]{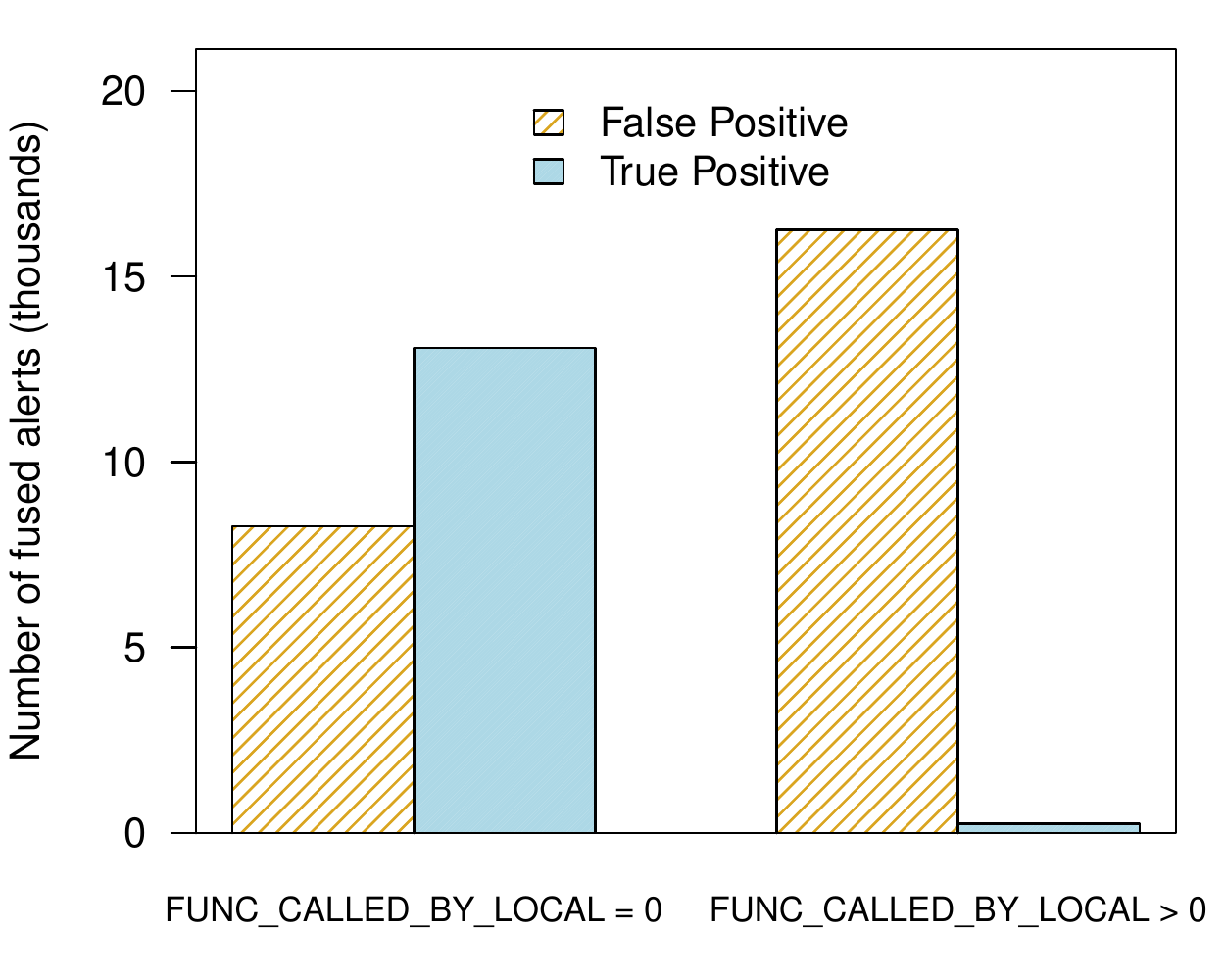}
  \vspace{-3ex}
  \caption{Labeled alerts vs. function caller counts}
  \label{fig:count.of.callers.vs.count.alerts.true.false}
\end{figure}

The feature with the highest information gain is the field \\
\ttt{FUNC\_CALLED\_BY\_LOCAL}, from the CCSM metrics tool, representing the ``number of local functions calling this function''. Fig.~\ref{fig:count.of.callers.vs.count.alerts.true.false} shows that alerts are almost always true positives when the function that contains the alerted line of code is called by at least one local function.
%% As expected, the checkerID feature has high (second-highest) impact. A particular checker may be very precise (e.g., look for a particular code flow, data flow, and type flow) or might use an imprecise heuristic such as simply looking for a string match that is only correct sometimes, leading to detection variations discussed in~\cite{delaitre2013massive}. Similarly, it is not suprising to see the CWE feature is important (rated sixth in importance), but less highly so. Code flaw types vary in how precisely they can automatically be detected. As expected, the CWE feature is lesser in importance than the checkerID. CheckerIDs represent a more granular entity: one tool's analysis for one flaw. A CWE often describes a larger concept of a flaw type that multiple checkerIDs from the same tools may may to, even without that checkerID set completely checking for all types of instances of that CWE. Different checkerIDs mapped to a single CWE may have varying precisions, whether the checkerIDs are from the same FFSA tool or different ones. However, in general some CWE types can be detected more precisely than others, leading to its feature importance. 
%The CCSM \ttt{RAW\_KW\_LONG\_CNT} feature holds the metric ``Count of 'long' keyword (in raw source)''.

\subsection{Results using speculatively mapped alerts} \label{subsection:speculative.mapping.results}

We re-trained LightGBM -- our best classifier -- separately on each of the training data sets $AF_{speculative}(T, d)$ (as per Sections \ref{subsection:spec_mapping}, \ref{section:train.test}). Experimenting with setting various thresholds $T$ on the $d=$forward (or backward) match percentages is a way to include speculatively mapped alerts of varying levels of "speculativeness" in the training data for the classifiers. This section summarizes these alternative training data sets in terms of the performance of the resulting trained classifier on test set $AF_{test}$.

Figure \ref{fig:SM_training_value} shows performance metrics for LightGBM (our best classifier) at each matching threshold. The red (blue dashed) lines show the performance at each threshold applied to the backward (forward) match percentage. To understand whether inclusion of speculatively-mapped alerts in the training set improves the performance of the classifier, compare the red and blue lines against the black point at the right-hand side of each graph, which marks the performance of LightGBM trained on data that excludes all speculative mappings (i.e. a threshold just greater than 1). 

From Figure \ref{fig:SM_training_value}, we conclude that the addition of the speculative alerts did not substantially affect the performance of the classifiers. This was surprising in two ways, since we had (1) expected performance to be slightly better with (as opposed to without) the inclusion of high-threshold speculatively-mapped alerts and (2) expected performance to degrade with the inclusion of alerts at the lowest threshold, since many of the speculative mappings are presumably incorrect and might "poison" the training data.

Looking past the overall average classifier performance to the performance on individual CWEs, there is a possibility that speculative mappings may improve the classifier for individual CWEs that are otherwise severely underrepresented in the training data. At the most inclusive speculative mappings configuration (threshold $0$), there are 
%DO NOT EDIT DIRECTLY! Autoproduced in /zkurtz/classify/speculative_mappings/rc224_comparisons.R
102\unskip\space
CWEs in the training data, significantly more than the 
%DO NOT EDIT DIRECTLY! Autoproduced in /zkurtz/classify/speculative_mappings/rc224_comparisons.R
82\unskip\space
CWEs in the non-speculative data. 

We observed, however, that the classifier accuracy for well-represented CWEs was not significantly higher than the accuracy on underrepresented ones. For instance, the mean accuracy for the top
%DO NOT EDIT DIRECTLY! Autoproduced in /zkurtz/classify/ophelia/publish/compile_results_for_icse2019.R
20\unskip\space
most-common CWEs in our training data was 
%DO NOT EDIT DIRECTLY! Autoproduced in /zkurtz/classify/ophelia/publish/compile_results_for_icse2019.R
0.946\unskip, 
only slightly greater than the accuracy of 
%DO NOT EDIT DIRECTLY! Autoproduced in /zkurtz/classify/ophelia/publish/compile_results_for_icse2019.R
0.942\unskip\space 
obtained for the 
%DO NOT EDIT DIRECTLY! Autoproduced in /zkurtz/classify/ophelia/publish/compile_results_for_icse2019.R
20\unskip\space 
least common CWEs. This suggests that there is not a strong relationship between the number of fused alerts in the training data for CWEs of a particular type and the resulting classifier performance for that CWE. That is, it appears that signals that the classifier learns are not particularly CWE-specific, such that the classifier may do well even for new alerts corresponding to CWEs that were not represented in the training data.

% The figure fig:SM_training_value subsumes these previous tables:
% Table~\ref{table:speculative_100} ($100\%$), 
% Table~\ref{table:speculative_75}  ($75\%$),
% and Table~\ref{table:speculative_50}  ($50\%$).
% 
% \input{tables/speculative_100.tex}
% \input{tables/speculative_75.tex}
% \input{tables/speculative_50.tex}

\section{Data}
We have published the open-source data used to develop classifiers and make mappings in this work, within the ``RC\_Data'' dataset \cite{flynn:rcdata.data} That dataset includes Juliet Java test suite data that is not part of this paper's case study, but that data was developed using the auto-labeling system discussed in this paper.

\section{Conclusions, Limits, Future Work}
We developed a novel method that uses test suites to automatically generate a large quantity of labeled data for SA alert classifier development.  We implemented this in a software system and then in a case study, we generated a large quantity of labeled data for many different conditions, using the Juliet test suite. Initial tests of the resulting classifiers on partitioned alerts from the test suite data show high accuracy for a large number of code flaw types.
We developed an automated method to map checkerIDs to code flaw taxonomy conditions specified in test suites. We also created a mised automated-and-manual method that can greatly reduce mapping effort compared to fully-manual mapping while being more correct than fully-automated mappings.

With incorporation of test suites that cover more conditions (e.g., more CWEs), we expect that the methods demonstrated here would lead to successful classifiers for those additional conditions.
In general, the method and software system developed can be used with any FFSA and code metrics tools to rapidly develop labeled data archives with standard additions to the system: Incorporating a new tool requires a parser and uploads to the database, while any new FFSA tool requires checkerID mappings. The general method can use additional artifact characteristics (source code repository metrics, bug database metrics, and dynamic analysis metrics), but the software system we developed would need to be extended.

The Juliet test suite programs are small (generally consisting of 1-3 short files) and are synthetically constructed. Our future work includes testing the classifiers on natural codebases, including both widely-used publicly-available cost-free open-source codebases and non-public proprietary codebases belonging to collaborators. Also, we have started to include test suites like STONESOUP~\cite{iarpa:stonesoup.phase.3} with much larger test programs (in the case of STONESOUP, widely-used natural test programs injected with flaws) that also have more complex control, data, and type flows than the Juliet test suite.

\bibliographystyle{plain}
\bibliography{rapidclass-bibliography}

\end{document}